
\input phyzzx

\font\zfont = cmss10 scaled \magstep1
\def\ZZ{\hbox{\zfont Z\kern-.4emZ}}
\def\Slacc{\centerline{\it Stanford Linear Accelerator Center}
\centerline{\it Stanford University, Stanford, CA 94309} }
\def\SCIPP{\centerline {\it Santa Cruz Institute for Particle Physics}
\centerline {\it University of California, Santa Cruz, CA 95064}}
\def\ie{{\it i.e.}}
\def\eg{{\it e.g.}}
\def\etc{{\it etc.}}

\overfullrule 0pt
\rightline{SCIPP 93/05}
\rightline{SLAC-PUB-6090}
\vskip 0.60in
\date{March 1993}
\titlepage
\title{{Supersymmetry and the Nelson-Barr Mechanism}
\foot{Work supported in part by the U.S. Department of Energy.}}
\author{Michael Dine and Robert G. Leigh}
\address{\SCIPP}
\vskip.2cm
\author{Alex Kagan}
\address{\Slacc}
\vskip.5cm
\vbox{
\centerline{\bf Abstract}
One possible solution to the strong CP problem is that CP is an exact
symmetry, spontaneously broken at some scale.  Some years ago,
Nelson and Barr suggested a mechanism for obtaining $\theta=0$
at tree level in this framework, and showed that radiative
corrections were small in some non-supersymmetric models.
Further investigations suggested that the same could be true
in supersymmetric theories.  In this note, we show that such
solutions assume extraordinarily high degrees of degeneracy
among squark masses and among other supersymmetry breaking parameters.
We argue, using naturalness as well as expectations from string
theory, that this is not very plausible.
}

\submit{Physical Review D}
\endpage

\parskip 0pt
\parindent 25pt
\overfullrule=0pt
\baselineskip=18pt
\tolerance 3500
\endpage
\pagenumber=1
\chapter{Introduction}

\REF\muzero{H. Georgi and I. McArthur, Harvard
University Report No. HUTP-81/A011 (unpublished);
D.B. Kaplan and A.V. Manohar, {\it Phys. Rev. Lett.} {\bf
56}, 2004 (1986); K. Choi, C.W. Kim and W.K. Sze, {\it Phys. Rev. Lett.}
{\bf 61}, 794 (1988); J. Donoghue and D. Wyler, {\it Phys. Rev.} {\bf D45}
(1992) 892; K. Choi, Nucl. Phys. {\bf B383} (1992) 58.}
\REF\pq{R.D. Peccei and H.R. Quinn, {\it Phys. Rev. Lett.}
{\bf 38} (11977) 1440; {\it Phys. Rev.} {\bf D16} (1977)
1791.}
\REF\axion{S. Weinberg, {\it Phys. Rev. Lett.} {\bf 40}
(1978) 223; F. Wilczek, {\it Phys. Rev. Lett.} {\bf 40}
(1978) 279.}
\REF\axioncosmology{J. Preskill, M. Wise and F. Wilczek,
{\it Phys. Lett.} {\bf 120B} (1983) 127; L. Abbott and P. Sikivie,
{\it Phys. Lett.} {\bf 120B} (1983) 133; M. Dine and W. Fischler,
{\it Phys. Lett.} {\bf 120B} (1983) 137.}
\REF\nelson{A. Nelson, {\it Phys. Lett.} {\bf 136B} (1984) 387.}
\REF\barr{S.M. Barr, {\it Phys. Rev. Lett.} {\bf 53} (1984) 329.}
\REF\babu{K.S. Babu and R.N. Mohapatra, {\it Phy. Rev.} {\bf D41} (1990)
1286.}
\REF\carlson{E.D. Carlson and M.Y. Wang, Harvard University preprint
1992, HUTP-92-A057.}
Since the strong CP problem was first recognized,
three solutions have been suggested.
The first scenario has no observable $\theta$-parameter,
because $m_u=0$.\refmark{\muzero}
The second gives $\theta=0$ dynamically, as the result of the existence
of a Peccei-Quinn symmetry\refmark{\pq} and its resulting
axion.\refmark{\axion}
The third scenario posits that CP is
an exact symmetry of the microscopic equations, which is
spontaneously broken, in such a way that the effective
$\theta$ turns out to be small.\refmark{\nelson,\barr,\babu,\carlson}
To date, there is no
definitive experimental evidence on any of these possibilities.
In first order chiral perturbation theory, one finds that $m_u \ne 0$.
However, there is the possibility that higher order corrections may
invalidate this conclusion, and there has been much debate
in the literature about this possibility.\refmark{\muzero}
The axion solution
is tightly constrained by astrophysical
and cosmological considerations.\refmark{\axioncosmology}
Within an order of magnitude or so, the lower limit on the axion
decay constant, $f_a$, seems unassailable.  The cosmological limit
relies on assumptions about very early cosmology, which might
not be correct, but which are extremely plausible.
Perhaps the
most promising suggestion for implementing spontaneous CP violation
with small $\theta$ is due to Nelson and Barr.\refmark{\nelson,\barr}
We will review this solution below.  Apart from predicting
that $\theta$ is small, however, it is not clear that there are any
generic low energy consequences of this picture; the resolution to
the strong CP problem is found among some massive states,
and the low energy theory is typically just a standard KM model.

\REF\banksdixon{T. Banks and L. Dixon, {\it Nucl. Phys.}
{\bf B307} (1988) 93.}
\REF\banksdine{T. Banks and M. Dine, {\it Phys. Rev.} {\bf D45} (1992)
424.}

In the absence of any definitive experimental answer, it is important
to look more closely at how plausible each of these solutions
may be from a purely theoretical perspective.  Necessarily this
requires making certain assumptions.  It seems reasonable to assume,
first, that any fundamental theory should not contain exact global
continuous symmetries.  This idea finds support both from string theory,
where the only continuous symmetries are gauge symmetries,\refmark{
\banksdixon} and from
considerations of the effects of quantum gravity on low-energy field theories.
Discrete symmetries are, however, quite plausible.  These do arise
in string theory, where in general they appear to be gauge symmetries,
and thus protected from violation by wormholes or other phenomena.
{}From this viewpoint, discrete symmetries are almost certainly necessary
if one is to implement the axion solution and might also be necessary
for the $m_u =0$ solution to
the strong CP problem.  Consider first $m_u=0$.  If we assume, for
the moment, that there is no new physics between $m_W$ and $m_{Pl}$,
all that is required is to suppress one Higgs coupling, and this is easily
achieved with a discrete symmetry.\foot{Note that this symmetry will
suffer from an anomaly.  However, in string theory, discrete symmetries
with anomalies arise frequently.  In the known examples, the
discrete anomalies are identical for all groups.\refmark{\banksdine}
In the present case,
this might imply the masslessness of some neutrino(s), for example.}
Higher dimension operators will make an extremely tiny contribution
to, \eg, $d_n$. Alternatively, in models in which the quarks obtain
their mass from mixing with
\REF\seesaw{A. Davidson and K.C. Wali, {\it Phys. Rev. Lett.} {\bf 59} (1987)
393; S. Rajpoot, {\it Phys. Lett.} {\bf 191B} (1987) 122; D. Chang and R.N.
Mohapatra, {\it Phys. Rev. Lett.} {\bf 58} (1987) 1600.}
heavy vector-like isosinglet quarks,\refmark{\seesaw} a massless up
quark could arise if there are fewer pairs of isosinglet up quarks than
down quarks. In this case there is no need for a discrete symmetry which
distinguishes it from the other quarks.
\REF\wittenpq{E. Witten, {\it Phys. Lett.} {\bf 149B} (1984) 351.}
\REF\pqdiscrete{G. Lazarides, C. Panagiotakopoulos and Q. Shafi,
{\it Phys. Rev. Lett.} {\bf 56} (1986) 432;
J. Casas and G. Ross, {\it Phys. Lett.} {\bf 192B} (1987) 119;
M. Dine, SCIPP-92-27.}

If we cannot impose global symmetries, Peccei-Quinn symmetries
must be accidental consequences of gauge and discrete symmetries.
In string theory, such Peccei-Quinn symmetries arise
automatically, but the decay constants of the associated axions are
of order $m_{Pl}$.\refmark{\wittenpq}
It is conceivable, as noted above, that this may be acceptable.  If we insist,
however, that the axion decay constant be of order $10^{10}-10^{11}$ GeV,
something further is required.  In Refs. \pqdiscrete, it was shown that,
within the framework of models with low energy supersymmetry,
discrete
symmetries can, in principle, lead to such a symmetry.  It is necessary,
however, to suppress operators up to very high dimension, and this
seems to require rather intricate patterns of discrete symmetry \dash\
certainly much more intricate than might be required for $m_u=0$.

\REF\intscale{M. Dine, V. Kaplunovsky, M. Mangano, C. Nappi and N.
Seiberg, {\it Nucl. Phys.} {\bf B259} (1985) 549.}
\REF\branco{L. Bento, G.C. Branco and P.A. Parada, {\it Phys. Lett.} {\bf
267B} (1991) 95.}

Finally, we turn to the real subject of this paper:  the plausibility
of the Nelson-Barr mechanism.  The basic idea is to arrange, by a
judicious choice of representations and symmetries, that
the tree level quark mass matrix, while complex, has real
determinant.  For definiteness, we will focus in this
paper on (supersymmetric)
models in which, beyond the usual particles of the
(minimal supersymmetric) standard model
there are some additional isosinglet
quarks with charge $\pm {1/3}$, denoted $q$ and $\bar q$.
These particles gain mass of order some scale $\mu$.
In addition, there are some standard model singlet particles,
${\cal N}^i$.  The superpotential in the $d$-quark sector is assumed
to take the form:
$$W_{d}= \lambda_{d,ij} Q^i \bar d^j H_1 + \mu q \bar q +
\gamma_{ij} q{\cal N}^i \bar d^j \eqn\genericmodel$$
The parameter $\mu$ may itself be the expectation
value of some scalar field. The singlet vev's and $\mu$ are typically
of order some large intermediate or GUT mass scale.  It is assumed that
the vev's of the ${\cal N}$ fields spontaneously break CP, while
$\mu$ is real. Then the fermion mass matrix takes the form:
$$m_F=\left(\matrix{A & B \cr 0 & \mu}\right)\eqn\nelsonmatrix$$
where the matrices $A$ and $\mu$ are real, while $B$ is complex.
This matrix has $arg~det~m_F=0$ and so the quark masses do not
contribute to $\theta$ at tree-level. The above is, essentially, the
supersymmetrized version of the minimal Nelson-Barr
model.\refmark{\branco}  As we will explain later, one framework
for obtaining a mass matrix of this type is provided by $E_6$
theories, such as those which appear in superstring theories.

\REF\barrmasiero{S.M. Barr and A. Masiero, {\it Phys. Rev.}
{\bf D38} (1988) 366.}
\REF\thooft{G. 't Hooft, in {\it Recent Developments in Gauge Theories},
G. 't Hooft et. al., Eds., Plenum (New York) 1980.}

In all models of this type, however,
there is no symmetry that explains the absence of $\theta$ at tree
level, and so one must ask what is the effect of radiative corrections.
In non-supersymmetric theories, Nelson\refmark{\nelson}
showed that radiative corrections can be sufficiently small to
give $\theta<10^{-9}$. The supersymmetric case was studied by
Barr and Masiero,\refmark{\barrmasiero} who
argued, again, that radiative corrections were small.  The main point
of the present paper is that the latter analysis, while essentially
correct, relied on a set of strong assumptions, which cannot be expected
to hold in general:  one must require an extremely high level
of degeneracy among the squark masses, and one must also demand
a very tight proportionality between certain pieces of the squark
and quark mass matrices.  We will argue that these conditions
are quite difficult to satisfy; indeed, the limits from
$K^o$-$\bar K^o$ mixing are already quite problematic in these models,
and as we will see the limits from $\theta$ are orders of magnitude
stronger. We are left with the view that the $m_u=0$ or axion solutions
of the strong CP problem are the most plausible presently known, at
least within the framework of supersymmetry.

\REF\segre{S. Barr and G. Segre, preprint BA-93-01 (1993).}
\REF\vadimjan{L. Ibanez and D. Lust, {\it Nucl. Phys.} {\bf B382} (1992)
305; B. de Carlos, J.A. Casas and C. Munoz, {\it Phys. Lett.} {\bf 299B}
(1993) 234; V.S. Kaplunovsky and J. Louis, CERN-TH-6809/93, UTTG-05-93.}

In the rest of this paper, we will elaborate upon these ideas.
In the next section, we will briefly review the limits on degeneracy
and proportionality which arise from the neutral kaon system, and
then describe the potential one loop contributions to $\theta$ which
give rise to the various constraints. In the third section, we
review some general aspects of soft-breaking terms in supersymmetric
theories. We explain why degeneracy is not expected to hold, in
general. We use 't Hooft's naturalness criterion to argue for limits
on degeneracy and proportionality.\refmark{\thooft} Indeed,
recent results in string-inspired models\refmark{\vadimjan}
support these limits. In Section 4, we enumerate the requirements
on the theory imposed by $K^o$-$\bar K^o$ and $\theta$ in a generic
Nelson-Barr model. We will see that they are
quite severe.  Some of these conditions are analyzed in more
detail in Section 5.  We will distinguish here two types of models.
The most promising are a class of models considered recently
by Barr and Segre,\refmark{\segre} in which large scales are generated
by heavy ${\cal N}$-type fields which
are much more massive than the susy-breaking scale, $m_{susy}$.
Even for these, however, one needs to make a set of strong assumptions,
which are not natural (in the sense of 't Hooft\refmark{\thooft}).
The other type of models we discuss are those in which large
intermediate mass scales are generated by light ${\cal N}$-type fields
with masses of order $m_{susy}$, as is typical in string-inspired
models.  We will argue that it is highly unlikely
that these constraints can be satisfied in this instance either. In our
conclusions, we will comment on possible ``ways out."

\chapter{One Loop Contributions to $K^o$-$\bar K^o$ and $\theta$}

\REF\masiero{M. Dugan, B. Grinstein and L. Hall, {\sl Nucl. Phys.} {\bf
B255} (1985) 413;
F. Gabbiani and A. Masiero, {\sl Nucl. Phys.} {\bf B322}
(1989) 235.}
It is well known that the neutral kaon system requires
that the squark mass spectrum exhibit a high degree of degeneracy.
The strongest limits come from box diagrams with the
exchange of gluinos.\refmark{\masiero}  From the contribution
of these diagrams to the real part, the most stringent limit obtained
is roughly of the form
$${\delta m_{Q}^2 \over m_{Q}^2}{\delta m_{\bar d}^2
\over m_{\bar d}^2}{1 \over m_{susy}^2} \lsim 10^{-10}
{\rm GeV}^{-2}\eqn\realkaon$$
whereas from the imaginary part one obtains a limit (with a phase) two
orders of magnitude stronger.  There are also similar bounds involving
only left-handed or right-handed squarks which are about an order of
magnitude weaker.  The quantities $\delta m_{Q}^2$, $\delta m_{\bar
d}^2$ are corrections to the left-handed and right handed degenerate
down squark mass matrices, $m_Q^2{\times \bf 1}$ and $m_{\bar
d}^2{\times \bf 1}$, respectively, and $m_{susy}^2$ is
typically of order the squark or gluino masses.\foot{In the quark mass
eigenstate basis it is the $ds$
entries which enter into the above constraints, but in the absence of a
detailed theory of flavor it is natural to take all $\delta m^2$
entries of same order.}
These limits are already quite striking. However, as we will
see, in models with spontaneous CP violation, one loop
corrections to $\theta$ give even stronger constraints on squark
degeneracy.

In conventional model-building, as in the minimal supersymmetric
standard model, one usually assumes a very simple structure for
the soft breaking contributions to the potential for the light
fields.  First, for the
terms of the type $\phi^* \phi$, where $\phi$ is some scalar field,
one takes (using the same symbol for the scalar component of
a multiplet as for the multiplet itself):
$$V_{\phi\phi^*}= m^2\sum_i \left(\vert Q_i\vert^2 + \vert \bar u_i\vert^2
+ \vert\bar d_i\vert^2\right) .\eqn\minimalassumption$$
Clearly this is a strong -- perhaps one should say drastic -- assumption.
It is certainly violated by radiative corrections.  Moreover, it is not
enforced, in general, by any symmetry and one does not expect
it to hold, in general, at tree level in any fundamental theory (it is
not the case in string theory, for example\refmark{\vadimjan}).
One makes a similar, drastic assumption for the cubic terms in the
potential:  one assumes that they are exactly proportional to the
superpotential,\foot{We are treating the
superpotential here as a homogeneous polynomial.  More general
cases will be considered below.}
$$V_{3}=A W(\phi) + {\rm h.c}. \eqn\aparameter$$
Again, this will not be respected by
radiative corrections and will not hold, generically, at tree
level.  In the next section, we will  consider, in the framework
of hidden-sector supergravity models, what these assumptions
mean.  For now, we simply note that they must hold to a rather
good approximation in order to avoid unacceptable flavor-changing
neutral currents.  Indeed, from the kaon system, the
limits we mentioned above imply that
these ``degeneracy" and ``proportionality"
conditions must hold to a part in $10^2$ or $10^4$, depending
on the value of the supersymmetry-breaking scale and
the nature of CP violation.  In order
that $d_n$ be small enough, one has restrictions on the
gluino and $A$-parameter phases as well.

In supersymmetric models of spontaneous CP violation where one has arranged
vanishing of $\theta$ at tree-level, there are a variety of
possible contributions to $\theta$ at one-loop
order.  These are given by
$$\delta \theta=Im \Tr\;\left[ m_u^{-1} \delta m_u + m_d^{-1}
\delta m_d\right]-3
{\rm Arg} \;{\tilde m}_3~. \eqn\delth$$
$\delta m_{u,d}$ represent the one-loop corrections to the tree level
quark mass matrices, $m_{u,d}$, and ${\tilde m}_3$ is the gluino mass including
one-loop contributions. In order to analyze these quantities, we first need to
make a few
more stipulations about the underlying model. To obtain
vanishing $\theta$ at tree level the tree level gluino mass
must be real (a non-zero phase represents a
contribution to $\theta$). All terms in
the Higgs potential must also be real.
To accomplish this we assume that supersymmetry breaking dynamics do not
spontaneously break CP, so that at some large scale the theory is
completely CP-invariant and, in particular, all supersymmetry breaking
terms are real.

\FIG\gluinograph{One loop diagram which can contribute a phase
to the gluino mass.  Matter fields may be light or heavy.}
\FIG\oneloop{One loop diagrams which will contribute a phase to
the quark mass matrix, if strict degeneracy or proportionality do
not hold.}
Nelson-Barr models will contain both light and heavy intermediate or
GUT scale quark fields.  Let us focus on the light fields and
denote the corrections to degeneracy by
$$\delta V_{\phi\phi^*} = Q^* \delta \tilde m_Q^2  Q +
\bar{d}\delta \tilde m_{\bar d}^2 \bar d^*
+\bar{u}\delta \tilde m_{\bar u}^2{\bar u}^* \eqn\deltaphistar$$
and proportionality by
$$\delta V_3= Q \delta A_d \bar d H_1 + Q \delta A_u\bar u H_2 . \eqn\deltaa$$
In general, integrating out the heavy fields leads to contributions to
some of the above terms {\it{which will be complex}} due to spontaneous
CP violation.
Bounds on degeneracy from $K^o$-$\bar K^o$ have been given above.
As noted, there are similar bounds on proportionality, the
most stringent one for the real part given by \refmark{\masiero}
$${\delta A_d <H_1 > \over m_{susy}^3} \lsim 10^{-5}
{\rm GeV}^{-1},\eqn\proprealkaon$$
while the limit for the imaginary part again is two orders of magnitude
stronger.

At the one-loop level there are a variety of contributions to $\theta$,
indicated in Figs. 1 and 2, in which the terms of
eqs. \deltaphistar\ and \deltaa\ appear as mass insertions.
Fig. 1 is the one-loop correction to the gluino mass.  This will
be complex, for example, if the corrections to the $A$ parameters in eq.
\deltaa\ are complex.  To satisfy the bound on $\theta$, this graph will
require that a certain relation hold for the soft-breaking
masses of heavy fields to about a part in $10^7$.

Fig. \oneloop\ corresponds to corrections
to $\theta$ coming from corrections to the
quark masses.  In the limit of exact degeneracy and proportionality,
these graphs yield contributions to $\theta$
proportional to $\Tr m_f m_f^{-1}$, which are clearly real.
($m_f$ here denotes the light fermion mass matrices).  Insertions of
$\delta V_{\phi\phi^*}$ and
$\delta V_3$ can yield complex corrections.  One dangerous possibility
arises from two insertions of  $\delta V_{\phi\phi^*}$.
This leads to a correction to the $d$-quark
mass which is in general complex, and proportional to $m_b$ and off-diagonal
terms in $\delta \tilde m_{\bar Q}^2$ and $\delta \tilde m_{\bar d}^2$.
These terms will therefore have to satisfy stringent limits.
A simple calculation gives:
$${{\left(\delta \tilde m^2_Q\right)\left(\delta
\tilde m^2_{\bar d}\right)} \over m_{susy}^4} < 10^{-9}
\eqn\bigconstraint$$
where $m_{susy}$, again, refers to some typical supersymmetry-breaking
mass.  The limit is so strong because a factor $O(m_b / m_d)$ appears in
the correction to $\theta$.

One might imagine that one would get extremely strong limits on
proportionality from this graph as well; in general it will give a contribution
to $\theta$ of order $\alpha_s /\pi\;  {\rm Im}~\Tr~ \delta A_d m_f^{-1}$.
It turns out, however, that in Nelson-Barr models this
trace is real. Stringent limits do
arise on products of $\delta A$ and $\delta \tilde m^2$
type terms, \eg,
$${(\delta A_d)\;(\delta \tilde m^2_{\bar d})
 \over m_{susy}^3}< 10^{-9} .
\eqn\anotherbigconstraint$$

These are striking constraints, which, in general, are considerably
stronger then the $K^o$-$\bar K^o$ limits discussed above.
In the rest of this paper
we will ask whether they might plausibly be satisfied.

\chapter{Some general aspects of soft-breaking terms in supersymmetric
theories.}

In this section, we review some aspects of supersymmetric theories,
and explain the origin of the non-degenerate and non-proportional terms.
For definiteness, we consider the case of $N=1$ supergravity theories,
with supersymmetry broken in a hidden sector.  Such a theory is described,
in general, by three functions, the K\"ahler potential, $K$, the
superpotential $W$, and a function $f$ which describes the gauge
couplings.  Here we focus on the form of the scalar potential.  Defining
a metric on the space of fields by
$$g_{i \bar j}= {\partial^2 K \over \partial \phi^i \partial
\phi^{j*}}\eqn\kahlermetric$$
and defining also
$$d_i = {\partial K \over \partial \phi^{i}}$$
the general potential is
$$V= e^{K}[({\partial W \over \partial \phi^i} +
d_i W)g^{i \bar j}
({\partial W \over \partial \phi^j} + d_{j} W)^* - 3 \vert W \vert^2]
. \eqn\generalpotential$$

\REF\hlw{L. Hall, J. Lykken and S. Weinberg, {\it Phys. Rev.} {\bf D27}
(1983) 2359.}
Our assumption that supersymmetry is broken in a hidden sector
means that there are two sets of fields: $z_i$, responsible for
supersymmetry breaking, and the ``visible sector fields," $y_i$;
$$W=g(y) + h(z).\eqn\visibleandhidden$$
The scales are such that
$$({\partial h \over \partial z^i} + {\partial K \over \partial z^i} h)
\sim m_{susy}m_{Pl}.\eqn\hidden$$
As discussed long ago by Hall, Lykken and Weinberg,\refmark{\hlw}
universality is the assumption that there is an approximate $U(n)$ symmetry
of the K\"ahler potential, where $n$ is the number of chiral multiplets
in the theory.  Frequently one takes simply
$$K= \sum \phi_i^* \phi_{ i} .\eqn\flatmetric$$
Clearly, this is an extremely strong assumption.  The Yukawa couplings
of the theory exhibit no such symmetry.  It does
not hold, for example, for a generic superstring
compactification, where the symmetry violations are
simply ${\cal O}(1)$.\foot{For a review of the relevant issues, see
Ref. \vadimjan .}  One can try to invent scenarios to explain
some approximate flavor symmetry. However, as we will discuss below,
't Hooft's naturalness condition suggests a
limit on how successful any such program can be.

We can characterize the violations of universality quite precisely.
For small $y$, we can expand $K$ in powers of $y$.  Rescaling
the fields, we can write\foot{Requiring that the $y_i$'s be canonically
normalized gives a condition on $<\ell_{ij}>$. But $\ell_{ij}\neq 0$.}
$$K= k(z, z^*)+ y_i y_i^* +  \ell_{ij}(z,z^*) y_i y_j^*
+ h_{ij} (z,z^*) y_i y_j + ....\eqn\kahler$$
There is no reason, in general, why $\ell_{ij}$ should
be proportional to the unit matrix, so
the $zz^*$ components of the metric will contain
terms involving $y_i y_j^* $ which are non-universal.
Plugging into eq. \generalpotential\ yields non-universal
mass terms for the visible sector fields.  In general, there
is no symmetry which can forbid these couplings.  For example,
$\ell = z^* z$ cannot be eliminated by symmetries.

Violations of proportionality arise in a similar manner.  The term
$\ell_{ij}$ in the K\"ahler potential leads to off-diagonal terms in the
$z^*y_i$ terms in the metric.  These in turn, from eq. \generalpotential,
lead to non-universal corrections to the cubic couplings.
One might imagine forbidding the dangerous terms by symmetries.
In particular, if no hidden-sector fields have Planck scale vev's (as
in models of gluino condensation)
and if couplings of the type $z y^* y$ are forbidden
by symmetries, then these corrections to the metric would be suppressed.
However, any such symmetry would also forbid a coupling
of the hidden sector fields to the gauge fields, required in order
to obtain a gaugino mass.

Recently, it has been found\refmark{\vadimjan} that string
theories give a soft supersymmetry-violating sector with degenerate
scalars at tree level if the dilaton F-terms dominate over moduli. This
is typically not the case when the most important non-perturbative
effect is gaugino condensation. However, even if the dilaton does dominate,
one expects universality to break down at order $\alpha_{str}/\pi$.

\REF\dkl{M. Dine, A. Kagan and R.G. Leigh, SCIPP-93-04 (1993).}
So we see that neither degeneracy nor proportionality are assumptions
which we would expect to hold generically.
The most obvious way to explain such features is to suppose
that there is an underlying flavor symmetry.
Any such symmetry must be approximate (it may arise,
for example, from a gauge symmetry broken at some energy
scale; see Ref. \dkl\ for a recent effort to build such models).
Even without considering a specific theory with such a flavor
symmetry, we can apply 't Hooft's naturalness criterion\refmark{
\thooft} to assess the plausibility of a given degree of degeneracy or
proportionality. In particular, we need to ask whether
the theory acquires greater symmetry as we assume that some
condition on masses holds.  Thus, for example, we  don't expect
degeneracy or proportionality between color-neutral scalars and
color-triplet squarks
to hold to better than $\alpha_s /\pi$, or degeneracy between
fields with different $SU_L(2)$ quantum numbers to hold
to better than $\alpha_W /\pi$.  Similarly, we don't expect
degeneracy for different flavors with the same
gauge quantum numbers to hold to better than
powers of Yukawa couplings.

\chapter{One loop corrections to $\theta$ in a Generic Model}

The supersymmetric Nelson-Barr model we'll consider is
the  minimal one described in the introduction:
in addition to the usual quark and
lepton families, we have an additional pair of isosinglet down quark
fields, $q$ and $\bar q$, as well as some
singlet fields, ${\cal N}_i$ and $\bar {\cal N}_i$.
It is straightforward to consider models with several $q$ and $\bar q$
fields, and with additional types of singlets.
The terms in the superpotential which give rise to the quark mass
matrix are
$$W= \mu q \bar q + \gamma^{ij} {\cal N}_i q \bar d_j
+ H_1\lambda_{ij} Q_i \bar d_j\eqn\basicw$$
(the terms in the superpotential involving $u$ quarks and leptons
will not be important for our considerations). $Q \bar q$ terms can be
forbidden via either gauged $U(1)$ or discrete symmetries.
\REF\frampton{P.H. Frampton and T.W. Kephart, {\it Phys. Rev. Lett.}
{\bf 65} (1990) 1549.}

One framework for obtaining a mass matrix of this kind is suggested
by $E_6$ models, such as those which appear in
superstring theories.
\foot{Some of the remarks here have
appeared earlier in work of Frampton and
 Kephart.\refmark{\frampton}}
In these models, generations of quarks and
leptons arise from the {\bf 27} representation.  Under $O(10)$, the
{\bf 27} decomposes as a {\bf 16}, a {\bf 10} and a {\bf 1}.  The
{\bf 16} contains
an ordinary generation of quarks and leptons,
plus a field which we denote by ${\cal N}$;
the {\bf 10} contains an additional $SU(2)$-singlet quark and antiquark
pair, $q$ and $\bar q$. We denote the $O(10)$-singlet by $S$.
Suppose that $E_6$ is broken at a high
scale down to a rank-6 subgroup, such as $SU(3) \times SU(2) \times
U(1)_Y \times U(1)_a \times U(1)_b$.
Then the dimension-four terms in the superpotential include the
couplings
$$W_{d}= Q \lambda \bar d^j H_1 + \lambda_S S q \bar q +
q{\cal N} \gamma \bar d\eqn\genericmodel$$
(here we have adopted a matrix notation for the various Yukawa
couplings; in general
there will be several $S$, ${\cal N}$, $q$ and $\bar q$
fields).  If the fields $S$ have real vev's, while the ${\cal N}$
fields have $CP$-violating, complex vev's, the mass matrix has the
structure of eq. \nelsonmatrix.  Moreover, plausible mechanisms
have been suggested for obtaining such vev's.  In particular, in
``intermediate scale scenarios," it has been noted that
the $S$ and ${\cal N}$ fields can readily obtain vev's of order
$m_I= \sqrt{m_{3/2} m_{Pl}}$.\refmark{\intscale}
One can easily check that for a finite range of parameters, the
${\cal N}$ vev's can be complex while the $S$ vev's are real.
In such a model, it is necessary to forbid a variety of other
couplings if one is to avoid a tree-level contribution to
$\theta$ and to meet other phenomenological requirements.
This can be done using discrete symmetries.
We will not present an example here, however, since,
as we will see, loop corrections almost inevitably
lead to serious problems.

Before estimating $\theta$, it is important to first consider
the questions of degeneracy and proportionality in models
of this kind.  We will assume that $\mu$ is some large
scale, such as $10^{11}$  GeV.  Our concern, then, is whether or not
the {\it light} squark mass matrix exhibits degeneracy and proportionality.
This requires that we integrate out the fields with mass of order
$\mu$.  Let us first examine the theory at scale $\mu$.
Because the soft-breaking terms are much
smaller than $\mu$, it is helpful to consider what the
theory looks like in their absence.  For both the
left-handed ($q$, $Q$) and right-handed ($\bar q$, $\bar d)$
sectors there is one state with
a mass of order $\mu$, and three light states.
For the left-handed states, the massive state is simply
$q$; for the right-handed states, it is
$$\bar D = ({1 \over m_D^2 + \mu^2})^{1/2}(M_Da^i \bar d_i + \mu \bar q)
\eqn\heavyguy$$
where
$$M_D^2=\vert \gamma^{ji}{\cal N}_j \vert^2\eqn\massMD$$
and
$$a^i ={1 \over M_D} \gamma^{ji}{\cal N}_j\eqn\vectora$$
Note that we have defined $\vec a$ so that $\vec a^{\dagger} \vec a=1$.

The relevant soft breaking terms at this scale are of two types.
Using the same letter
for the scalar component of a supermultiplet
as for the multiplet itself, the $\phi\phi^*$ type terms are of the form
$$V_1= Q^* \tilde m_Q^2  Q +
 {\bar{d}}  \tilde m_{\bar d}^2  {\bar{d}}^*+
+  \tilde m_q^2 q q^*
+ \tilde m_{\bar q}^2 {\bar{q}}{\bar{q}}^*\eqn\softmass$$
Here, $\tilde m_{\bar d}^2$,  $\tilde m_{Q}^2$, \etc, are
assumed to be of order $m_{susy}^2$.
The $\phi\phi$ and $\phi\phi\phi$ terms are of the form:
$$V_{2}=  Q A_d  \bar d H_1 + A_{\mu} \mu q \bar q
+ {\cal N} A_{\gamma} q \bar d\eqn\softtri$$
We have defined the $A$'s so that they are dimensionful quantities
of order $m_{susy}$.  $A_d$ and $A_\gamma$ are matrices.  Note our
definition of $A_d$ differs from that of the previous section,
where $A_d$ was defined on the {\it light states only}.

As we'll see, near degeneracy of the full $4 \times 4$ right-handed
down squark mass matrix and of $\tilde m_Q^2$ will be required, so we write
$$\tilde m_{\bar d}^2 = \tilde m_{\bar d}^2 \times {\bf 1}+\delta
m_{\bar d}^2 ~ ; ~~
\tilde m_{\bar q}^2 = \tilde m_{\bar d}^2 + \delta
\tilde m_{\bar q}^2~ ; ~~\tilde m_Q^2 = \tilde m_Q^2 \times {\bf 1} +\delta
\tilde m_Q^2.
\eqn\fourbyfourdegen$$
Near proportionality of $A_d$ is also necessary and we write
$$A_d \equiv A_d {\lambda_d  } + \delta A_d .\eqn\threebythreeprop$$
(Again, the notation $\lambda_d$ is being used in a different sense
than previously.)

Let us now examine the form of the various mass matrices.  Calling
$m_d=\lambda_d H_1$, the fermion mass matrix and
 its inverse have the form
$$m_F= \left( \matrix{m_d & M_D \vec a \cr 0 &\mu} \right)
{}~~~~~m_F^{-1}= \left( \matrix{m_d^{-1} & {M_D \over
\mu} m_d^{-1}  \vec a \cr 0 &\mu^{-1}} \right) .
\eqn\fermionmatrix$$
The matrix $m_F$ has the Nelson-Barr form, and its determinant is real.

In view of our remarks in Section 2, it is the form of
the scalar mass matrices which particularly concerns us.
Consider first the $\phi \phi^*$ type terms.  For the
squarks in the {\bf 3} representation of $SU(3)$
, these take the form, on the full $4\times 4$
set of states:
$${\cal M}_{LL}^2 =
 \left(\matrix{\tilde m_d^2 + m_d^T m_d &
M_D m_d^T \vec a \cr \vec a^{\dagger}m_d M_D &
\tilde m_q^2 + M_D^2 + \mu^2}\right).\eqn\dsquark$$
Similarly, for the $\bar{\bf 3}$ squarks (``right-handed"
squarks), we have:
$${\cal M}_{RR}^2 =
 \left(\matrix{\tilde m_{\bar d}^2 + m_d m_d^T + M_D^2
\vec a \vec a^{\dagger} &
\mu M_D \vec a \cr \mu M_D \vec a^{\dagger}
& \tilde m_{\bar q}^2 + \mu^2}\right) .\eqn\dbarsquarks$$
Finally, for the $\phi\phi$-type matrix, which connects the $3$
and $\bar 3$ squarks, we have
$${\cal M}_{RL} ^2 = \left(\matrix{A_d <H_1 > + \mu_H{H_2\over H_1} m_d
& M_5^2 \vec b \cr 0 & A_{\mu} \mu} \right)
\eqn\lrmatrix$$
where $\mu_H$ is the coefficient of the $H_1H_2$ term in
the superpotential, and $M_5$ and $\vec b$ are defined by
$$M_5^2 b^i= A_\gamma^{ji}{\cal N}_j + \left(
{\partial W \over \partial {\cal N}_j}\right)^*
\gamma^{ji}~~ ; ~~~\vec b^{\dagger} \vec b=1\eqn\bdefinition$$
Note, in general, $\vec a$ is not proportional to $\vec b$ (see Section 5).

We are now in a position to integrate out the heavy fields to
obtain an effective lagrangian for the light fields.  The first
question we should address is that of degeneracy and proportionality.
Even before considering $\theta$, degeneracy is necessary to understand
the properties of neutral kaons, and proportionality is required
to suppress other contributions to $d_n$.  To construct the mass matrices for
the light fields and examine these questions, it is convenient to introduce
projection operators onto the light and heavy states, in the supersymmetric
limit (\ie, ignoring corrections of order $m_{susy}$ or $<H>$).
The projector onto the heavy ``left-handed" states is just
$${\bf P}_L^h= \left( \matrix{0 & 0 & 0 & 0 \cr 0& 0&0&0 \cr 0 & 0 &0 & 0 \cr
0& 0 & 0 & 1}\right) .\eqn\pqlight$$
The projector onto the light states is simply:
$${\bf P}_L^l= {\bf 1}-{\bf P}_L^h .\eqn\pqlight$$
The projectors onto the right-handed states are slightly more complicated.
Onto the massive state, it is
$${\bf P}_R^h=({1 \over M_D^2 + \mu^2})\left( \matrix{M_D^2 \vec a
\vec a^{\dagger} & \mu M_D \vec a \cr \mu M_D
\vec a^{\dagger} & \mu^2}\right) .
\eqn\pdbarh$$
The light projector is then just ${\bf P}_R^l={\bf 1}-{\bf P}_R^h$.

\FIG\intout{Simple diagram which describes integrating out the
massive fields to obtain corrections to the squark mass matrices.}
With these we can immediately project the fermion mass matrix
onto the light states:
$$m_F^l= {\bf P}_R^l m_F {\bf P}_L^l = \left(
\matrix{({\bf 1}-{M_D^2 \over M_D^2 + \mu^2}\;\vec a \vec a^{\dagger}) m_d& 0
\cr  -{\mu M_D  \over M_D^2 + \mu^2}\; \vec a^{\dagger} m_d& 0}\right).
\eqn\lightf$$
To work out the scalar matrices requires somewhat greater care.
In addition to performing a projection of the type described above,
there are couplings of heavy to light states.  Taking these into
account, and integrating out the heavy states, as in Fig. \intout,
yields additional corrections to the light squark mass matrices.
Consider, first, the result of projecting the matrices onto the light
states.  For the left-handed squarks, one has
$${\cal M}_{LL}^{l~2}= \left( \matrix{\tilde m_d^2 + m_d^T m_d & 0
\cr 0 & 0 } \right) \eqn\lightdsquark$$
If $\tilde m_d^2$ is proportional to the unit matrix, ignoring the
terms proportional to fermion masses, this expression is proportional to
${\bf P}_L^l$, \ie, for degeneracy in the left-handed sector, it is sufficient
that the $3 \times 3$ matrix $\tilde m_d^2$ be degenerate.

The situation for the right-handed squarks is more complicated.
Here one has
$${\cal M}_{RR}^{l~2}= {\bf P}_R^l  {\cal M}_{RR}^2 {\bf P}_R^l .
\eqn\lightdbarsquark$$
Rather than write down this expression in detail, let us simply
consider a particular case as an example; suppose in eq. (4.7)
we set $\delta \tilde m_{\bar d}^2 =0$.
Then, ignoring terms proportional to quark masses,
$${\cal M}_{RR}^{l~2} = \tilde m_{\bar d}^2 {\bf P}_R^l
+ \delta \tilde m_{\bar q}^2 \;{M_D^2\over (M_D^2+\mu^2)^2}\;\left(
 \matrix{\mu^2 \vec a \vec a^{\dagger} & -\mu M_D \vec a \cr
-\mu M_D \vec a^{\dagger} & M_D^2 }\right) . \eqn\almostdegen$$
$\delta m_{\bar q}^2 \approx 0$, then, is required in order to obtain
degeneracy, \ie,
we need $4 \times 4$ degeneracy if the light $\bar d$-squark
matrix is to be degenerate. In terms of the naturalness criteria
we developed in the preceeding section, this means that the $\bar q$
and $\bar d$ fields should have the same gauge quantum numbers;
this will certainly not be the case in models which use gauge
symmetries to obtain the Nelson-Barr structure.  Moreover, all of the
Yukawa couplings involving the $\bar q$ fields must be small.

This is not the whole story, however, since, in the basis we have chosen, there
are in general couplings of the light squarks to the heavy squarks of order
$m_{susy} \mu$.  Thus integrating out heavy squarks will give additional
contributions to the light squark mass matrices of order $m_{susy}^2$.
A simple calculation shows that the only light-heavy couplings
of order $\mu$ arise from the matrix ${\cal M}_{RL}^2$
and involve the couplings of the massive $q$
to the light $\bar d$ fields.  The required coupling is described by
the matrix:
$$ {\cal M}^2_{lh}= {\bf P}_R^l {\cal M}_{RL}^2 {\bf P}_L^h
= \left( \matrix{0 &M_5^2({\bf 1}-{M_D^2 \over M_D^2 + \mu^2}
\vec a \vec a^{\dagger}) \vec b -{\mu^2 A_{\mu} M_D \over
M_D^2 + \mu^2} \vec a  \cr 0 & -{\mu M_D M_5^2 \over
M_D^2 + \mu^2} \vec a^{\dagger} \vec b + {\mu A_{\mu} M_D^2
\over M_D^2 + \mu^2}} \right) .
\eqn\leftrightcoupling$$
Note that this coupling vanishes if $\vec a = \vec b$ and $M_5^2
= A_{\mu} M_D$.  If these conditions do not hold,
integrating out the massive field then leads to a shift in
${\cal M}_{RR}^{l~2}$,
$$\delta {\cal M}_{R}^{l~2} ={\cal M}^2_{lh}{1 \over M_D^2 + \mu^2}
{\cal M}^{2~\dagger}_{lh}.
\eqn\intoutresult$$
The resulting expression is rather involved.  The main point, however,
is that the expression is not proportional to ${\bf P}_R^l$,
so these terms are not degenerate.  In fact, generically, the resulting
non-degeneracy would exceed the bounds from $K^o$-$\bar K^o$.  One can
attempt to suppress $\delta {\cal M}_{R}^{l~2}$
by a judicious choice of parameters.  For example, for
$\vec b=\vec a$ and $\mu \ll M_D$, these couplings are suppressed by
$\mu^2 /M_D^2$.  The ratio $\mu^2 /M_D^2$ can be sufficiently small,
\eg, of order $10^{-4}$, and still permit a realistic light quark
mass matrix.  A hierarchy of entries in $\vec b$ reflecting the
hierarchy of entries in $m_d$, \eg, ${b}_{1,2}\ll {b}_{3}$,
can also suppress the dangerous $sd$-entries in $\delta {\cal M}_{R}^{l~2}$.
As we will see shortly, there is also a suppression
for large $\mu/M_D$; however, in this limit, it is easy to see
that the KM phase is likewise suppressed.\foot{If the KM phase is
suppressed, one is lead to consider models where susy box diagrams
dominate $\epsilon$.}

In string-inspired models with low energy generation of intermediate
mass scales, the coupling of eq.
\leftrightcoupling\ does not vanish.
Recently, Barr and Segre\refmark{\segre} have argued that the
conditions $\vec a = \vec b$ and $M_5^2= A_{\mu} M_D$
will hold provided that the ${\cal N}$ fields are much more
massive than $m_{susy}$, given certain assumptions
about proportionality.   However, as we will see in the next
section,  these assumptions violate the naturalness criteria we
have set forth; one expects that this condition cannot hold to
better than $\alpha_s /\pi$.

In order to insure near proportionality for the light states to
satisfy the limits from $K^o$-$\bar K^o$ it is sufficient
that the $3\times 3$ block of ${\cal M}^2_{RL}$
be approximately proportional, as in eq. \threebythreeprop .
In this case, the light left-right squark mass matrix has the
form
$${\cal M}_{RL}^{l~2}= {\bf P}_R^l {\cal M}_{RL}^2 {\bf P}_L^l =
(A_d + \mu_H{H_2 \over H_1}) m_F^l + \delta A_d^l <H_1>
,\eqn\proportionalityholds$$
where $\delta A_d^l$ is obtained from eq. \lightf\ by replacing $m_d$
with $\delta A_d$.
So approximate proportionality holds for the light quark and squark
states.  Integrating out the heavy fields at
tree level yields only contributions suppressed by powers of
$m_{susy}/ \mu$.

So, even before worrying about $\theta$, we see that in supersymmetrized
Nelson-Barr models there is the potential for severe violations of degeneracy,
even making the assumptions typical of supergravity models.
In addition to assuming that the various $3 \times 3$ parts
of the squark mass matrices are nearly degenerate and proportional at the
high energy scale, we need that the $\bar d$-squark mass matrix have a
$4 \times 4$ degeneracy, to about a part in $10^3$, and that dangerous
entries in the
matrix $\delta {\cal M}_R^{l~2}$ be very small.
These conditions are quite disturbing.  Even if there is
exact degeneracy of the $4 \times 4$ squark mass matrix at
tree level, loop corrections will violate it.  These will typically
be enhanced by large logarithms.
In order that one-loop corrections not be too large, the Yukawa couplings,
$\gamma^{ij}$, must be small, about $10^{-2}$ or smaller.
If the $\bar q$ and $\bar d$ fields carry different gauge quantum numbers,
as would be typical in simple string models, one expects corrections of
order ${\alpha\over\pi} ln(m^2_{Pl}/\mu^2)$.
In the following, we will simply assume that this
condition is somehow satisfied, \eg, through small couplings
and suitable quantum numbers.  But already we view this as
unattractive.

Let us now ask how severe are the constraints arising from the smallness
of $\theta$.  First consider the gluino mass diagram.
This receives contributions not only from light states
but from heavy states as well.  The result is easy to work
out in terms of the projectors above, and it is complex
in general if $\vec a \ne\vec  b$:
$${\rm Im}~ \delta m_{\lambda}\sim
 {\alpha_s \over 4 \pi}
{M_D^2 \over (M_D^2+\mu^2)} {M_5^2\over M_D}
{\rm Im}~ \vec a^{\dagger} \vec b.\eqn\imlambda$$
Indeed, this diagram leads to the requirement that the phases of these
vectors line up to about one part in $10^{-7}$.  Certainly the simplest
way to satisfy this is $\vec b =\vec a$; we will assume
this to be the case in the remainder of this section.
In the next section, we will investigate this condition and argue that
it is not natural.
The light fermion contributions to the gluino mass
lead to a  weaker limit on proportionality.  If $A_d$ is not
proportional to the unit matrix, one will obtain a complex
result, in general.  This will give a limit suppressed by
powers of the $b$-
quark mass over the susy-breaking scale:
$${{\rm Im}~{\vec a}^{\dagger}\delta A_d  m_d^{T}{\vec a}
\over m_{susy}^3} \;{<H_1> M_D^2 \over{ M_D^2 +\mu^2}} \lsim 10^{-7}.
     \eqn\gluinoalimit$$

More significant limits arise from the graphs of Fig. 2.  From
one proportionality violating insertion and one degeneracy
violating insertion we obtain:
$${{\rm Im}~{\vec a}^{\dagger}  \delta A_d \lambda_d^{-1}
\left(\delta\tilde m^2_{\bar d}-
\delta \tilde m^2_{\bar q}\times
{\bf 1}\right) \vec a  \over m^3_{susy}}\;{M_D^2 \over {M_D^2 + \mu ^2}}
\lsim 10^{-6},\eqn\deltaadeltam$$
and
$${{\rm Im}~ {\vec a}^{\dagger} \delta A_d \lambda_d^{-1} {\vec a}\over
m_{susy}^3}\; {(M_5^2 -A_{\mu} M_D)^2 \mu^2 \over{ (M_D^2 + \mu^2)^2}}
\lsim 10^{-6}. \eqn\deltaadeltamlh$$
The second equation is essentially due to the contribution of
$\delta {\cal M}_{R}^{l~2}$ , \ie, of integrating out heavy fields,
to right-handed squark non-degeneracy.

{}From Fig. 2, with two degeneracy-violating
insertions, we find contributions to $\theta$ of order:
$$ 10^{-1}{\alpha_s\over4\pi}\;{A_d {\rm Im}~\vec a^{\dagger} m_d \delta\tilde
m_Q^2
m_d^{-1} \left(\delta \tilde m^2_{\bar d}-
\delta\tilde m^2_{\bar q}\times{\bf 1}\right)
\vec a  \over m_{susy}^5} {M_D^2 \over {M_D^2 +
\mu^2}},\eqn\twodeginsertions $$
which lead to the constraints
$${\left( \delta \tilde m_{\bar q}^2,
 \delta \tilde m_{\bar d}^2\right)\over m_{susy}^2}
{(\delta \tilde m_Q^2) \over m_{susy}^2} \lsim
10^{-9}.\eqn\finallimits$$
Non-degeneracy due to $\delta {\cal
M}_{R}^{l~2}$ leads to a contribution of order
$$ 10^{-1} {\alpha_s\over4\pi}\;{ {\rm Im}~\vec a^{\dagger} m_d \delta\tilde
m_d^2 m_d^{-1}\vec a  \over m_{susy}^3
}{{(M_5^2 -A_{\mu} M_D)M_D} \over {M_D^2 + \mu^2}}.
 \eqn\twodeginsertionslh $$

How plausible is it that one can satisfy these constraints?
The gluino diagram constraint is satisfied by $\vec b=\vec a$;
we will explore in the next section the meaning of this condition.
The degeneracy and proportionality constraints require satisfying
limits on $\delta \tilde m_d^2$,
$\delta \tilde m_{\bar d}^2$, $\delta
\tilde m_{\bar q}^2$ and $\delta A_d$, which are considerably more
stringent than those obtained from $K^o$-$\bar K^o$ mixing.  It is
hard to comprehend how they could be satisfied in the absence of a
detailed theory of flavor. These constraints require as well
a condition on $M_5$, or perhaps some other condition on parameters.
For example, the equality $M_5^2=M_D A_{\mu}$ would eliminate
light-heavy squark couplings, so that contributions to $\theta$
which arise from integrating out heavy fields would vanish, see eqs.
\deltaadeltam\ and \finallimits. However, this condition, as we will
see in the next section, requires exact degeneracy (at the Planck
scale); again, naturalness arguments preclude this possibility.
Alternatively, one could try to
exploit the fact that $M_D \ll \mu$ would suppress all of the
above contributions to $\theta$. Unfortunately, this strategy is
limited by the fact that the induced KM phase or the phase entering
the susy box graph would be of order ${M_D^2 \over \mu^2}$ and so this
ratio must be $\gsim 10^{-2}$ in order to generate large enough $\epsilon$.

\chapter{The $\vec b= \vec a$ and $M_5^2=M_D A_{\mu}$
constraints}

We would now like to comment on whether or not one is likely to
find $\vec a = \vec b$ in a given model.
In light of some recent observations of Barr and
Segre,\refmark{\segre} we will distinguish
two cases:  one, suggested by  string-inspired models, where the
${\cal N}$ fields have masses of order $m_{susy}$
yet generate large intermediate mass scales by exploiting (approximate)
F-flat and D-flat directions, and one in which the ${\cal N}$ fields
have much larger mass.  The potential for trouble exists because
both $\phi\phi\phi$ soft breaking terms and F-terms contribute to
off-diagonal components of the squark mass matrix.
The calculation of Barr and Segre is particularly simple,
and we will describe it first.  Suppose that the fields, ${\cal N}$,
couple to some other fields, $Y$, and that all of these fields
have mass (at the minimum of the potential) much greater
than $m_{susy}$.  Motivated by supergravity models,
these authors suppose that the holomorphic soft breaking
terms in the potential have the form
$$V_{soft}= a~m_{susy} W + b ~
m_{susy} \sum \phi_i {\partial W \over \partial \phi_i}
+ {\rm h.c.} \eqn\barrsegresoft$$
Now, because the masses of the ${\cal N}$ and $Y$ fields are
assumed to be large, it follows that
$$m_{susy}{\partial W \over \partial \phi_i} \ll {\partial^2 W
\over \partial \phi^2} \eqn\derivativeordering$$
Similarly, in obtaining the minimum, one can neglect the
soft-breaking terms of the type $\vert {\cal N}\vert^2$, \etc\
As a result, at the minimum, the potential satisfies
$$m_{susy}b \phi_j{\partial^2 W \over \partial \phi_i \phi_j}
= -{\partial^2W \over \partial \phi_i \partial \phi_j}
{\partial W^* \over \partial \phi_j^*} \eqn\barrsegrecondition$$
But this gives immediately that
$${\partial W^* \over \partial \phi_i^*}= -b m_{susy}
\phi_i\eqn\partialequality$$
{}From this it follows from eq. \bdefinition\ that $\vec a = \vec b$.

Similarly, with these assumptions, it follows that $m_5^2 = A_{\mu}
M_D$.   To see this, note first that $A_{\mu}= a+2b$.
Also, $A_{\gamma}=a+3b$ in eq. \bdefinition.  So substituting
$\partialequality$ in eq. \bdefinition, we immediately obtain the
desired equality. It is crucial here that the parameter $b$ in
eq. \barrsegresoft\ is common to all terms in the superpotential.

However, from our earlier discussions of proportionality, we see
that this is not
natural in the sense of 't Hooft.  For example, the $a$ and $b$
parameters for the $q$ and ${\cal N}$ fields, \ie, for $A_{\gamma}$,
$A_{\mu}$ and for the purely singlet scalar potential, would be
expected to differ by amounts at least of order ${\alpha_s \over \pi}$,
since the ${\cal N}$ fields do not carry color.  This, in turn, leads
to violations of the $M_5^2=A_{\mu} M_D$
condition of this order.  This leads to unacceptably large
$\theta$; one is forced to try and suppress
$\theta$ by further choice of parameters (\eg, by taking $M_D/\mu $
very small, as discussed in the last section).  This prospect does
not appear to be feasible.

In models in which the ${\cal N}$ fields have masses of order
$m_{susy}$, the situation is no better.  In intermediate scale
models, in addition to the ${\cal N}$
fields, one adds $\bar{ \cal N}$ fields with opposite gauge quantum
numbers.  Intermediate mass scales are generated along the resulting
(approximate) F- and D-flat directions of the scalar potential.
As before, from eq. \bdefinition,
the condition $\vec a=\vec b$  will automatically
be satisfied if, at the minimum of the potential, we have:
$${\partial W\over \partial {\cal N}_a} = \alpha m_{susy} {\cal N}_a^*
\eqn\proport$$ for some (real) $\alpha$.
This condition is not completely implausible but we will see that it
relies on a certain form of the superpotential and soft breaking terms.
Denote the part of the
superpotential depending on the fields ${\cal N}$ and $\bar{\cal N}$
as $g({\cal N},\bar{\cal N})$.  In the string-inspired
intermediate scale scenario,
$g({\cal N},\bar{\cal N})$ is of dimension five, \eg, containing terms of
the form ${{\cal N}^2 {\bar {\cal N}^2}\over m_Pl}$.
Extremization of the potential gives:
$$
{\partial^2 g\over \partial {\cal N}_a\partial {\cal N}_b}
{\partial g^*\over \partial {\cal N}_b^*} +
{\partial^2 g\over \partial {\cal N}_a\partial \bar{\cal N}_b}
{\partial g^*\over \partial \bar{\cal N}_b^*} +
{\partial V_{SSB}\over \partial {\cal N}_a} + \sum_i
e_{\cal N}^{(i)} D^{(i)} {\cal N}_a^* = 0\eqn\minim$$
with similar equations for $\bar{\cal N}$.
(Here $V_{SSB}$ is the susy-breaking potential and the last term
arises from the $D^2$ terms in the potential, with $e_{\cal N}^{(i)}$
the cooresponding charges.)
All terms in this equation are of order $m_{susy}^2 m_I$.  Now if
$g({\cal N},\bar{\cal N})$ is homogeneous in ${\cal N}$ and $\bar{\cal N}$,
as is typical in intermediate scale scenarios, then eqs. \minim ,
supplemented by eq. \proport\ (with a similar equation for $\bar{\cal N}$,
with $\alpha$ replaced by $\beta$), are equivalent to
$$
(\alpha^2 +2\alpha\beta + A\alpha){\cal N}_a + \tilde{m}^2_{{\cal N},ab}
{\cal N}_b =0 \eqn\alpbet$$
where $\tilde{m}^2_{{\cal N},ab} $ is the soft-SUSY-breaking mass matrix
for ${\cal N}$. Unless $\tilde{m}^2_{{\cal N},ab} $ is proportional to
the unit matrix (\ie\ highly degenerate),
eqs. \proport\ and \alpbet\ are an overdetermined set of
equations for ${\cal N}_a$. Thus we see that, with assumptions on the
form of the superpotential, eq. \proport\ can be understood as a
degeneracy problem in the ${\cal N}$-sector.   By the
naturalness criteria we enunciated earlier, the level of degeneracy
we can expect in this sector is presumably of order the associated
Yukawa couplings.   Worse, in string-inspired models,
the $\bar q$, $\bar d$, and ${\cal N}$ fields typically
carry different quantum numbers under a gauged $U(1)$, and thus
the corresponding $A$ terms would be expected to differ by
amounts of order $\alpha$.

In models of this type, it seems even harder to understand the
$M_5^2=A_{\mu} M_D$ condition, even approximately.  Thus
again we are forced into some special region of parameters,
if these models are to be viable at all.

To conclude this discussion, we mention that one might hold out hope
that, given a complete theory of flavor,
sufficient degeneracy could be obtained. In fact, let us,
contrary to our naturalness arguments and stringy expectations,
assume exact degeneracy at the Planck scale. By studying the
renormalization group equations, one finds that most contributions
to $\theta$ may be kept small enough given sufficient suppression
of the couplings $\gamma_{ij}$. However, the contribution of eq.
\twodeginsertions\ is problematic if there is a gauged $U(1)$ which
distinguishes $\bar q$ and $\bar d$.\foot{This is commonly the
case in string-inspired models.} The (LL) part
receives a renormalization proportional to the up-quark Yukawa
couplings while the (RR) factor is renormalized by gauge
couplings. It is well-known that such renormalizations are not
problematic for FCNC's but the resulting $\theta$ in Nelson-Barr
models is much too large.

\chapter{Conclusions}
\REF\dynsusy{M. Dine and A. Nelson, SCIPP 93-01 (1993).}

We have found that the constraints on degeneracy and proportionality
of the squark mass matrices in Nelson-Barr scenarios are much
stronger than those obtained from data on flavor changing neutral
currents. Moreover, based on naturalness as well as expectations
from string theory, we have argued that these constraints are
not likely to be satisfied.

We are not able, of course, to say with certainty that a model
cannot be constructed which satisfies the constraints on degeneracy
between light scalar masses and proportionality of scalar versus
fermionic couplings of the light states which have been enumerated
here in a natural way.  We have remarked already that one can
construct models in which horizontal symmetries give rise to modest
squark degeneracies.  Perhaps more ingenious constructions can give
the higher degrees of degeneracy required. Those couplings which
might be expected (by 't Hooft criterion) to differ by $\alpha_s/\pi$
conceivably could be nearly equal in some unified framework.
Another possibility might be
scenarios incorporating dynamical supersymmetry breaking,
in which large degeneracies are expected since supersymmetry is broken
by gauge interactions which are flavor blind.\refmark{\dynsusy} It is
also interesting that in such schemes, scalar trilinear couplings are
expected to be greatly suppressed, which would help with the various
proportionality constraints, including $\vec b=\vec a$.

{}From all of this, we are left with the feeling that the
$m_u=0$ and axion solutions to the strong
$CP$ problem (in the framework of supersymmetry)
are the most plausible.  Successfully
implementing the Nelson-Barr scheme requires
a much more ingenious understanding of flavor physics
than has been offered to date.

\smallskip
\centerline{\bf Acknowledgements}
{We would like to thank A. Nelson for helpful conversations
about the Nelson-Barr mechanism, and T. Banks and N. Seiberg
for discussions of $m_u=0$ and other flavor issues.}

\refout
\figout
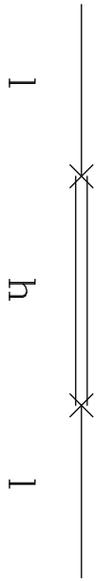
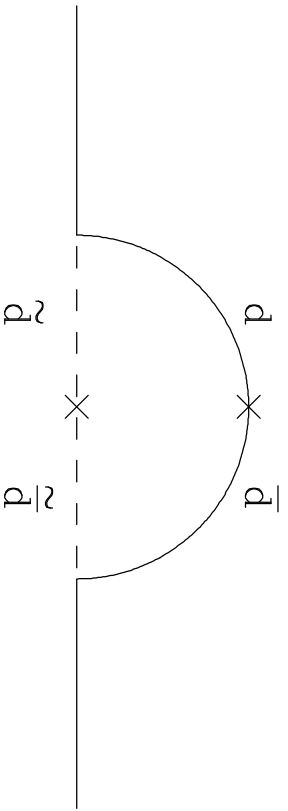
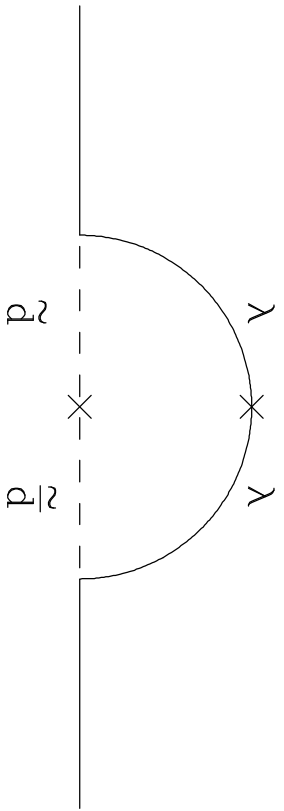
\end